\documentclass[twocolumn,pra,tighten,floatfix,showpacs, amssymb]{revtex4-1}
\usepackage{graphicx}

\def\<{\langle}
\def\>{\rangle}

\def\be{\begin{equation}}
\def\ee{\end{equation}}

\begin{document}
\preprint{cond-mat} \title{Entanglement Temperature and Perturbed AdS$_3$ Geometry}

\author{G. C. Levine and B. Caravan}

\address{Department of Physics and Astronomy, Hofstra University,
Hempstead, NY 11549}

\date{\today}

\begin{abstract}

Generalizing the first law of thermodynamics, the increase in entropy density $\delta S(x)$ of a conformal field theory (CFT) is proportional to the increase in energy density, $\delta E(x)$, of a subsystem divided by a spatially dependent entanglement temperature, $T_E(x)$, a fixed parameter determined by the geometry of the subsystem, crossing over to thermodynamic temperature at high temperatures. In this manuscript we derive a generalization of the thermodynamic Clausius relation, showing that deformations of the CFT by marginal operators are associated with spatial temperature variations, $\delta T_E(x)$, and spatial energy correlations play the role of specific heat.  Using AdS/CFT duality we develop a relationship between a perturbation in the local entanglement temperature of the CFT and the perturbation of the bulk AdS metric. In two dimensions, we demonstrate a method through which direct diagonalizations of the boundary quantum theory may be used to construct geometric perturbations of AdS$_3$.


\end{abstract}

\maketitle


Important connections have recently been forged between quantum information and gravitational physics.  A relationship between the change in entanglement entropy, $\delta S$, of a subregion of a conformal field theory (CFT) and an energy, $\delta E$, associated with an excited state of the CFT has been developed that generalizes the first law of thermodynamics \cite{MyersQFL}. Subsequently it was shown that the resulting generalized first law (GFL) was equivalent to the linearized Einstein Equations for perturbations of the AdS bulk \cite{RaamsQFL,TakaEE}. The GFL was further elucidated by Klich and coworkers \cite{KlichTe} who identified an "entanglement temperature" function by showing that the reduced density matrix of a subregion in a CFT could generally be written:
\begin{equation}
\label{Klich_dm}
\hat{\rho} = \frac{1}{Z}e^{-\int{dx\beta(x)\hat{T}(x)}}
\end{equation}
where $\hat{T}(x)$ is the energy density operator of the CFT and $\beta(x)$ is the inverse of the entanglement temperature that diverges at the boundary of the subsystem.  Within this framework, the GFL is simply derived by perturbing the ground state density matrix by $\delta \hat{\rho}$ and expanding the entropy, $S = -{\rm tr}\hat{\rho} \log{\hat{\rho}}$ to first order to yield:
\be
\label{QFL}
\delta S = \int{dx \beta(x) \delta T(x)}
\ee
where $\delta T(x) \equiv {\rm tr}(\delta \hat{\rho} \hat{T}(x)) $

In analogy to an equation of state, it is natural to ask how the entropy responds to a change in an external parameter such as an applied field. To see that some relation other than the GFL is needed we note that one might easily apply a field that increases the energy while decreasing the entanglement entropy.  A simple example, which we subsequently study, is a one dimensional noninteracting spinless fermion gas, corresponding to a $c=1$ CFT in the continuum. The marginal backscattering operator locally changes the energy density by a finite amount \cite{GNT}, however equation (\ref{QFL}) predicts a vanishing entropy correction if the operator acts at the boundary of the subregion where $\beta(x) \rightarrow 0$.  In contrast, backscattering is known to affect the entanglement entropy most strongly at the boundary, and weakly everywhere else \cite{EEimp}.  This suggests that the entanglement temperature {\sl itself} must be affected by an external field.

In this manuscript we use AdS/CFT duality to show that, for a $d=2$ CFT, a local marginal perturbation leads to a local perturbation of the entanglement temperature: $\beta(x) \rightarrow \beta(x) + \delta \beta(x)$.  Then, from equation (\ref{Klich_dm}), it follows that 
\be
\label{delta_entropy}
\delta S = - \int{dx dx^\prime \beta(x) \delta \beta(x^\prime) \langle \hat{T}(x)\hat{T}(x^\prime) \rangle}
\ee
where angle brackets denote expectation value with respect to the unperturbed density matrix.  Identifying the energy correlation with the specific heat, $c_V$, this equation is analogous to the Clausius relation of thermodynamics:
\be
\label{Clausius}
\delta S = c_V \frac{\delta T}{T}
\ee 
When $\beta(x)$ is associated with projected area of the minimal surface on the boundary of AdS, equation (\ref{delta_entropy}) will be shown to express a local relationship between the perturbed entanglement temperature and the perturbation to the AdS background metric.

We begin by describing the relationship between entanglement temperature and minimal surface area that appears in the AdS$_{d+1}$ construction for holographic entanglement entropy.  Given a minimal surface whose boundary defines a region $\Omega$ in the CFT is, the holographic entropy \cite{RyuTaka} may be computed from an integral over boundary coordinates:
\be
\label{area}
S =\frac{1}{4G} \int_\Omega{d^{d-1}x \sqrt{g}}
\ee 
The equation of the minimal surface spanning a radius $l$, $d-1$ dimensional ball of the CFT  is  $l^2 = z^2 + x^a x_a$, where $x^a$ ($a = 1,\ldots, d-1$) are the spatial boundary coordinates. Parameterizing the surface $\{\tilde{x}^{0}=z(x); \tilde{x}^{a} = x^a \}$, the induced metric is:
\be
\label{induced}
g_{ab} = G^0_{\mu \nu} \partial_a \tilde{x}^\mu  \partial_b \tilde{x}^\nu  = \frac{R^2}{z^2}(\frac{x^a x^b}{z^2} + \delta_{ab})
\ee 
where $G^0_{\mu \nu} = \eta_{\mu \nu}R^2/z^2$ ($\mu = 0,\ldots, d-1$) is the unperturbed spatial metric for a radius $R$ AdS$_{d+1}$ space. The determinant is $g = R^2 l^2/z^4$ and, defining the radial boundary coordinate $r^2 \equiv x^a x_a $,
\be
\label{beta}
\sqrt{g} = R\frac{l}{l^2 - r^2} = \frac{2 \pi R}{\beta(r)}
\ee
where $\beta(r)$ is the inverse of the entanglement temperature of a $d$-dimensional field theory, as defined by Klich \cite{KlichTe} for a radius $l$ subregion of a CFT on an infinite domain. Equation (\ref{beta}) expresses that entanglement temperature $T_{\rm E}(r) = \beta^{-1}(r)$ at a point $r$ on the boundary is associated with the projected differential area element of the minimal surface at the point associated with $r$. Using equation (\ref{beta}) and $c=3R/2G$, the holographic entropy formula (\ref{area}) is seen to be completely equivalent to integrating the entanglement temperature to find the entropy---equation I.9 of reference \cite{KlichTe}, written in integral form here for $d=2$,
\be
\label{KlichEntropy}
S = \frac{\pi c}{3} \int_{\Omega}{\frac{dx}{\beta(x)}}
\ee

We now consider the effect upon the entanglement temperature, $1/\beta(x)$, of applying an external field $\phi_0(x)$ coupled to an operator ${\cal{O}}(x)$ at a point $x$ in the boundary theory.  Specializing to $d=2$ the action of the boundary theory is deformed by:
\be
\label{deformation}
S^\prime = \int{d^2x \phi_0(x){ \cal{O}}(x)}
\ee
Following the AdS/CFT prescription, the applied external field, $\phi_0(x)$, is a source for a scalar field $\phi(z,x)$ in the gravity bulk, determined by the bulk-boundary propagator for AdS$_3$ ($B$ is a normalization factor):
\be
\label{b2b}
\phi(z,x,t) = B \int{dx^\prime dt^\prime \frac{z^\Delta}{(z^2 + (x-x^\prime)^2 - (t-t^\prime)^2)^\Delta} \phi_0(x^\prime, t^\prime)}
\ee
where $\Delta$ is the scaling dimension of ${\cal O}(x)$ and $B$ is a normalization constant.  We have also written space/time coordinates of the boundary theory explicitly. 

\begin{figure}[ht]
\includegraphics[width=6cm]{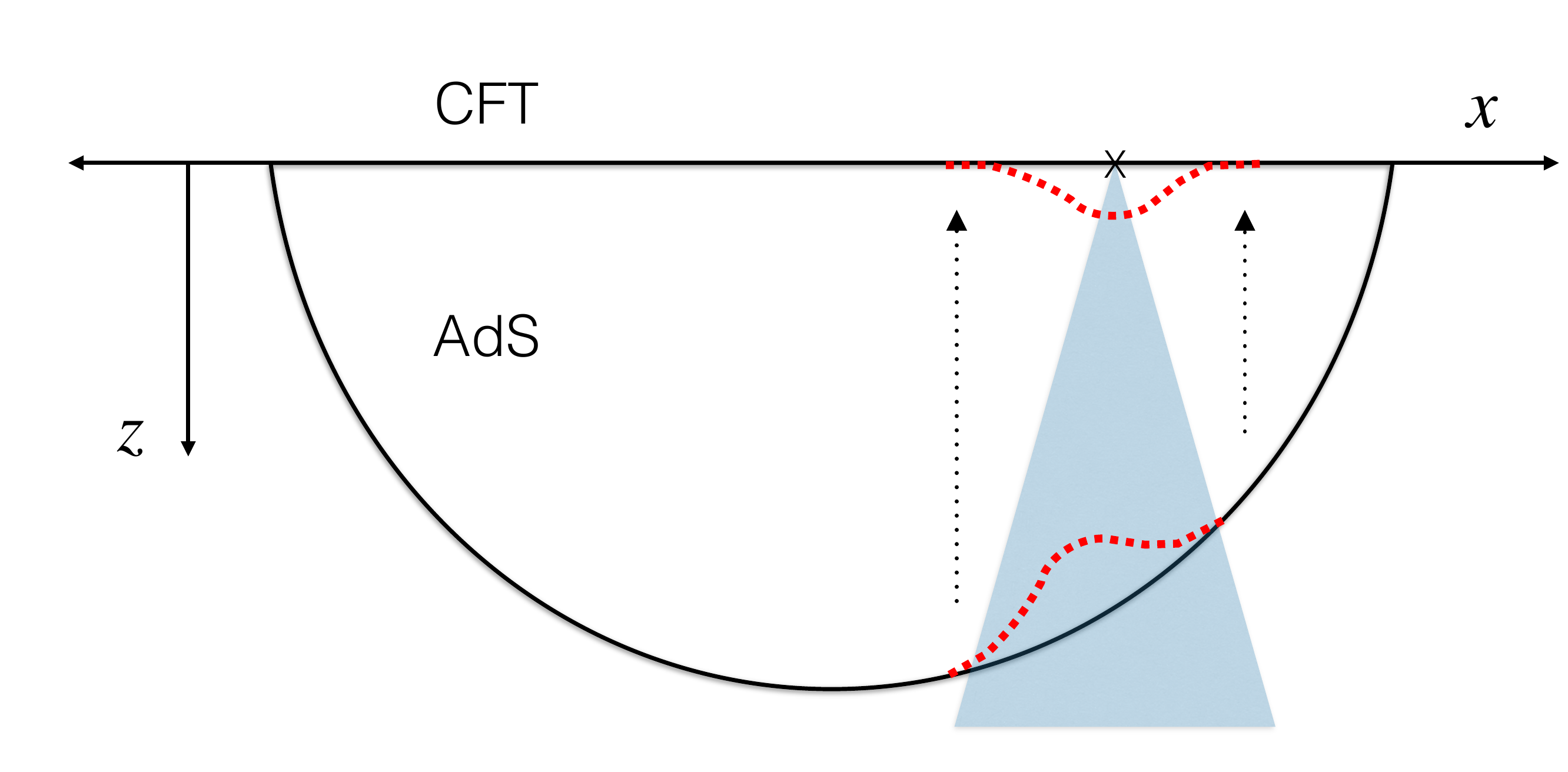}
\caption{Depiction of defect and holographic entropy construction of minimal surface. Localized, marginal scalar perturbation of CFT (defect marked by $\times$) propagates into bulk AdS (shaded region.) Bulk scalar field sources perturbation of metric over an extended region of minimal surface (dashes).  Perturbation to the minimal surface area projects back onto boundary, perturbing entanglement temperature over an extended region (dashes). }
\label{cartoon}
\end{figure}

$\phi(z,x)$ is a source for the Einstein equations governing perturbations to the AdS metric. These perturbations, in turn, induce local perturbations to the minimal surface area, thus perturbing $\beta(x)$ through equation (\ref{beta}).  As seen in the solution (\ref{b2b}), a point source at the boundary diffracts as the field propagates in the bulk, effecting a spatially extended perturbation of the entanglement temperature. This scheme is depicted in figure (\ref{cartoon}).

To confirm this heuristic analysis we have developed a procedure to numerically compute the spatial dependence of the entanglement temperature $1/\beta(x)$ in a lattice model of one dimensional free spinless fermions deformed by a marginal perturbation. In the continuum limit such a model realizes a two dimensional $c=1$ CFT.
\begin{figure}[ht]
\includegraphics[width=7cm]{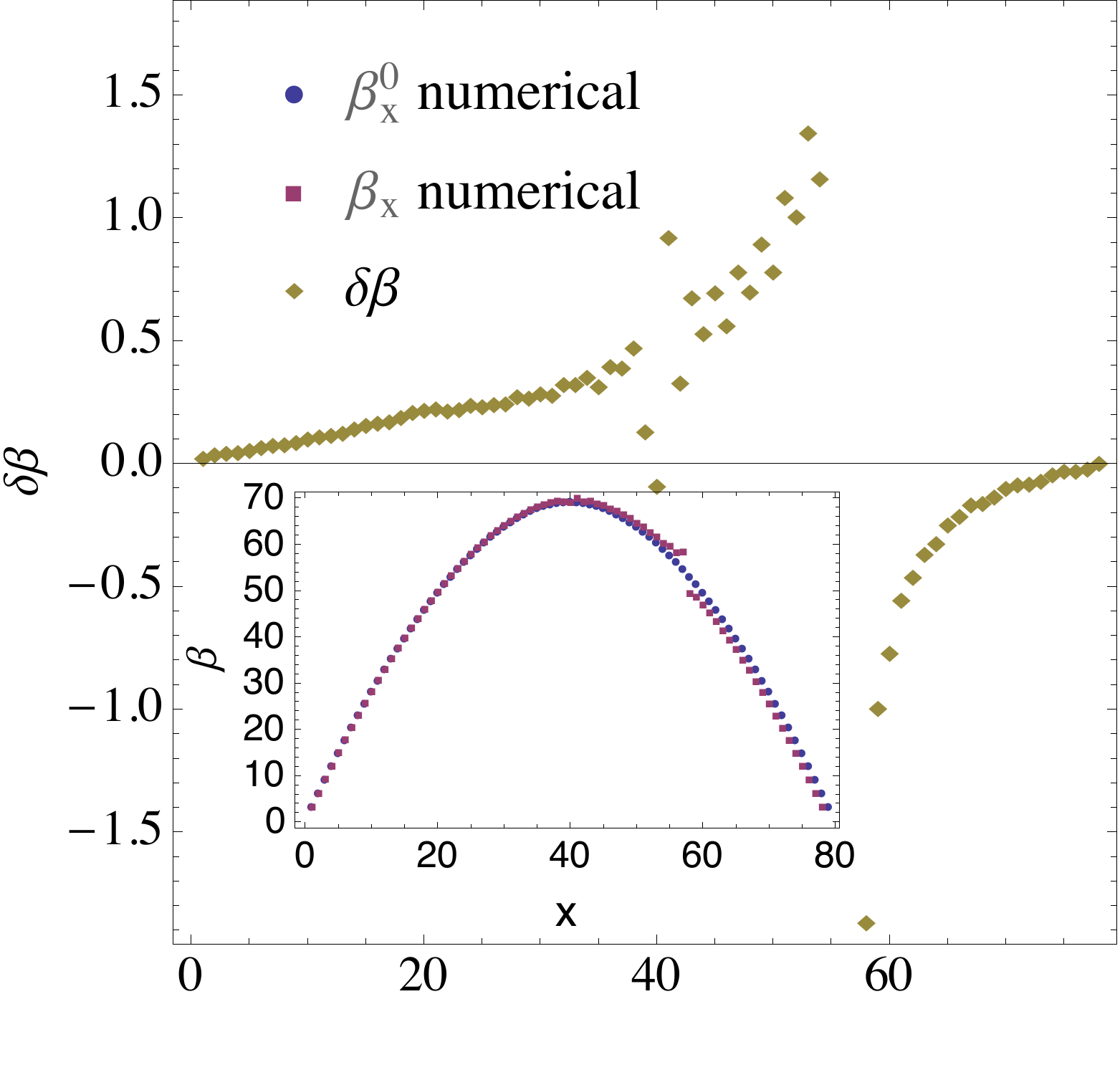}
\caption{Numerical calculation of lattice inverse entanglement temperature, $\beta_x$, of an $2l = 80$ site subregion on a $450$ site lattice with a defect $t^\prime = t/2$ at site $57$. Unperturbed $\beta^0_x$ (inset) is indistinguishable from analytic form $\beta_0(x) = \pi(l^2 -x^2)/l$ (not shown) on scale of figure. $\delta \beta_x = \beta_x - \beta^0_x$ shows the singularity at the defect position and algebraic decay ($\delta \beta_x \sim 1/x$) corresponding to diffraction of a point source in AdS space. }
\label{deltabetafig}
\end{figure}

Our basic numerical results are summarized in figure (\ref{deltabetafig}). $\beta^0_x$, computed without the perturbation, is nominally a parabola, corresponding closely to the discrete version of $\beta_0(x) = \pi(l^2 -x^2)/l$; the perturbation, $\delta \beta_x$, has odd parity about the defect position and decays algebraically corresponding to the spatially extended scheme described above.

We now turn to the calculation of entanglement entropy from $\delta \beta(x)$ to arrive at our central result, equation (\ref{delta_entropy}). Following that, we will work out the relationship between $\delta \beta(x)$ and the perturbation in AdS bulk metric that follows from the Ryu-Takaynagi formula for holographic entropy and gravitational perturbation theory.

Writing the perturbed reduced density matrix 
\begin{equation}
\label{perturbed_dm}
\hat{\rho} = \frac{1}{Z}e^{-\int_{-l}^l{(\beta_0(x) + \delta\beta(x))\hat{T}(x)dx}}
\end{equation}
with $Z$ set to normalize ${\rm tr} \hat{\rho} = 1$, we would like to expand the entropy $S = -{\rm tr}\hat{\rho} \log{\hat{\rho}}$ to first order in $\delta \beta(x)$. Let $A \equiv \int_{-l}^l{dx \beta_0(x) \hat{T}(x)} $ and $B \equiv \int_{-l}^l{dx \delta \beta(x) \hat{T}(x)} $ and consider an expansion in $\beta_0(x)$ keeping only ${\cal O}(\delta \beta(x))$ in the exponent:
\be
\label{expansion}
e^{A+B} = e^A e^B e^{-\frac{1}{2}[A,B]} e^{\frac{1}{6}[A,[A,B]]} \ldots
\ee
The operator $\hat{T}(x)$ satisfies the Virasoro algebra 
\be
\label{vira}
[\hat{T}(x),\hat{T}(x^\prime)] = D(x, x^\prime) \hat{T}(x) + i\frac{\pi}{3} c \partial_x^3\delta(x-x^\prime)
\ee
where $D(x, x^\prime) = 2 \pi i \delta(x-x^\prime) \partial_x - 4 \pi i \partial_x \delta(x-x^\prime) $ and the nested commutators are all linear in $\hat{T}(x)$. Using equation (\ref{expansion}),
\be
\label{nasty_exp}
 \hat{\rho} = \frac{1}{Z}e^{-\int{dx\beta_0(x)\hat{T}(x)}}\Big\{ 1-\int{\delta\beta(x_1)\hat{T}(x_1)} 
\ee
$$ - \frac{1}{2} \int{\beta_0(x_1)\delta\beta(x_2)D(x_1 x_2)\hat{T}(x_1) }$$
$$- \frac{1}{6} \int{\beta_0(x_1)\beta_0(x_2)\delta\beta(x_3)D(x_2 x_3)D(x_1 x_2)\hat{T}(x_1) } + \ldots \Big\} $$
where constant terms arising from the conformal anomaly (the last term in (\ref{vira})) have been suppressed.   Adopting the notation
\be
\langle \dots \rangle = {\rm tr}(\hat{\rho}_0 \ldots) = \frac{1}{Z_0}{\rm tr}(e^{-\int{\beta_0(x)\hat{T}(x)}}\ldots)
\ee
where $Z_0 \equiv {\rm tr}\exp{(-\int{\beta_0(x)\hat{T}(x)})}$, the expectation value of any operator $\langle \hat{O}(x) \rangle$  supported in the interval $[-l,l]$ yields its {\sl zero physical temperature}, infinite domain expectation value. Making the same expansion of $Z$ and noting that $\langle \hat{T}(x) \rangle = 0$, $Z$ may be replaced by $Z_0$ in equation (\ref{nasty_exp}). Equation (\ref{nasty_exp}) may now be viewed as a high temperature expansion (in $\beta_0(x)$) for each order of $\delta \beta(x)$. Similarly expanding the $\log{\hat{\rho}}$ factor appearing in the entropy definition,
\be
\label{entropy2}
S = \log{Z_0} - \int{dx dx^\prime \beta_0(x) \delta \beta(x^\prime) \langle T(x)T(x^\prime) \rangle}
\ee
to lowest order in $\beta_0(x)$ and $\delta \beta(x)$, establishing eq. (\ref{delta_entropy}).

 For a CFT expressing both chiral modes, the energy correlation function regularized by a spatial cut-off $\alpha$ is:
\be
\label{energy_corr}
\langle \hat{T}(x) \hat{T}({x^\prime}) \rangle = \frac{c}{2} [ \frac{1}{(x-x^\prime +i\alpha)^4} + \frac{1}{(x-x^\prime -i\alpha)^4}]
\ee
Performing the $x$ integration in (\ref{entropy2}),
\be
\label{deltaS_QM}
\delta S = - \frac{c}{3}\int{dx \frac{4 \pi l^2}{(l^2 - x^2)^2} \delta \beta(x)} 
\ee
Equation (\ref{deltaS_QM}) will be used to relate $\delta \beta(x)$ to the perturbed AdS$_3$ bulk metric. 

The comparable equation to (\ref{deltaS_QM}) on the gravity side relates the perturbation in holographic entropy to the perturbed bulk metric, computed from the back reaction of the bulk metric to the CFT perturbation (\ref{deformation}). At this point we also restrict our analysis to time-independent marginal perturbations ($\Delta = 1$).  The metric incorporating a perturbation $h_{\mu \nu}(z,x)$ ($\mu = 0, \ldots, d$) of pure AdS in the Fefferman-Graham gauge is:
\be
\label{metric}
ds^2 = \frac{R^2}{z^2}(dz^2 + (\eta_{\mu \nu} + h_{\mu \nu})dx^\mu dx^\nu)
\ee
The resulting entanglement entropy may be found by perturbing equation (\ref{area}) using the specific parameterization for the AdS$_3$ minimal surface spanning an interval $[-l,l]$ on the boundary: $x(\sigma) = l\sin(\sigma), z(\sigma) = l\cos(\sigma)$ for $\sigma \in [-\pi/2, \pi/2]$. Writing $\delta g$ in terms of $h_{xx}$, 
\be
\label{delta_holo_entropy}
\delta S = \frac{1}{8 G} \int{dx \frac{1}{\sqrt{g}}\delta g} = \frac{c}{12}\int{dx \frac{1}{l} h_{xx}(z(x),x)}
\ee
where we have made use of the relation between central charge, $c$, and properties of AdS: $c = 3R/2G$.  

Comparing equations (\ref{deltaS_QM}) and (\ref{delta_holo_entropy}), both expressing the central charge, $c$, suggests a relationship between perturbed entanglement temperature and the perturbed bulk metric. Consider a defect located at boundary coordinate $x=0$ and an asymmetric subregion $[\xi-l,\xi+l]$ centered at $x=\xi$. The perturbed entanglement temperature now depends upon $\xi$ for fixed $l$: $\delta \beta(x,\xi)$. The minimal surface may be regarded as a probe in that the metric perturbation $h_{xx}(x,z)$ at a point $(x,z)$ may be constructed  from a numerical computation of $\delta \beta(x,\xi(z))$,
where $\xi(z) = x- \sqrt{l^2 - z^2}$ describes a particular minimal surface passing through $(x,z)$. Then,
\be
\label{betatoh}
h_{xx}(x,z) = \frac{16 \pi l^3}{z^4} \delta \beta(x,\xi(z))
\ee

To compare our numerical computation of the perturbation of the lattice entanglement temperature, $\delta \beta_x$, to gravitational perturbation theory, $h_{xx}$ is computed for a localized applied field in the boundary theory $\phi_0(x^\prime, \omega) = \phi_0 \delta (x^\prime) \delta (\omega)$.  The scalar field in the bulk is found from equation (\ref{b2b}) to be:
\be
\label{bulk_phi}
\phi(z,x,t) =\phi_0  \frac{1}{\sqrt{2\pi}} \frac{z}{\sqrt{x^2 + z^2}}
\ee

This field now sources the Einstein equations through the reduced stress-energy tensor:
\be
\label{T00}
T_{tt} = -\frac{1}{4}[(\partial_z\phi)^2 + (\partial_x\phi)^2 ]
\ee

Following reference \cite{TakaEE}, $h_{xx} = h_{tt}$, and the linearized Einstein equation,
\be
\label{EE}
\partial_z h_{xx} - z\partial^2_z h_{xx} = 2z T_{tt}
\ee
may be integrated to yield the perturbed AdS$_3$ metric,
\be
\label{hxx}
h_{xx}(z,x) = B^{\prime} \frac{z^2}{4x^2}\log{\frac{z}{\sqrt{x^2 + z^2}}}
\ee
where the external field, $\phi_0$ and other constants have been absorbed into $B^\prime$.

In comparing $\delta \beta(x,\xi(z))$ and $h_{xx}(x,z)$, we focus on the important difference in the analytic structure describing the defect. Gravitational perturbation theory predicts a nonsingular behavior $\lim_{x \rightarrow p} h_{xx}(x,z) = B^\prime/4$ whereas our numerical calculations indicate a singular behavior:
 \be
 \label{approxbeta}
\delta \beta(x) \approx -a\frac{l^2}{x-p} {\rm sign }(p)
\ee

\begin{figure}[ht]
\includegraphics[width=6cm]{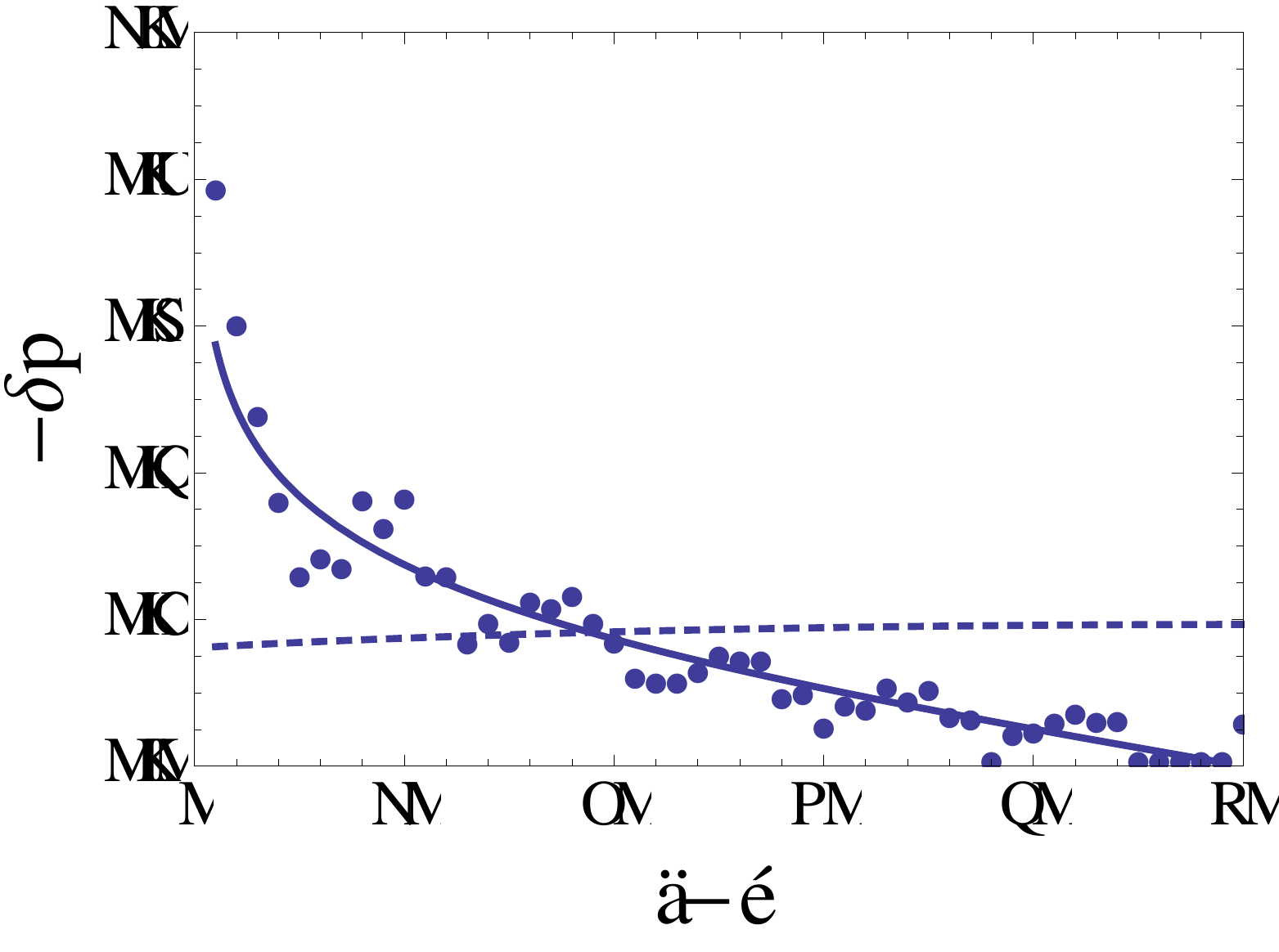}
\caption{Perturbation to entropy as a function of distance of defect from boundary ($l-p$). Symbols: exact numerical calculation on a $350$-site lattice with a subregion of size $2l=160$, $t^\prime =t/2$. Dashed curve: gravitational perturbation theory, equations (\ref{delta_holo_entropy}) and (\ref{hxx}) with $B^\prime = 1.6/l$. Solid curve: boundary theory calculation using equations (\ref{deltaS_QM}) and (\ref{approxbeta}) with $ a = 0.13$; see \cite{IntComment}. }
\label{deltaSDist}
\end{figure}

Numerically integrating equation (\ref{delta_holo_entropy}) using (\ref{hxx}) to determine the entropy leads to a nearly constant entropy as the defect is moved to the boundary (figure \ref{deltaSDist}). This behavior disagrees qualitatively with the exact computation of entropy shown in figure (\ref{deltaSDist}) which has a pronounced trend of large (negative) correction to the entropy as the defect approaches the boundary.  In contrast, the algebraic singularity (\ref{approxbeta}) when integrated in equation (\ref{deltaS_QM}) yields $\delta S \approx a\log{\frac{l-p}{l+p}}$, agreeing qualitatively with the exact entropy calculation as shown in figure (\ref{deltaSDist}) (see note \cite{IntComment}). The odd parity of $\delta \beta$---contrasting the even parity of $h_{xx}$---appears to be the correct qualitative behavior as well: a strong defect that isolates one smaller region of size $s$ has an entanglement temperature, $T_E$, that diverges as $s \rightarrow 0$. Therefore, $\delta \beta < 0$ on the side of the smaller region and $\delta \beta > 0$ on the side of the larger region. 

The disagreement between gravitational perturbation theory and entanglement calculations suggests that gravitational perturbation theory may miss a coordinate singularity associated with the exact dual of this deformed CFT. The singular behavior suggested by our calculation may bear some relation to the AdS/Schwarzschild solution used in \cite{TakaQuench} to study entanglement dynamics following a quench.

We briefly describe our numerical calculation.  Using slightly unconventional notation, our tight-binding Hamiltonian is 
\be
\label{ham}
H = \sum{\hat{T}_{x,y}} +\frac{t^\prime}{t}\hat{T}_{p,p+1}
\ee
where the kinetic energy density operator is written in terms of the single particle kinetic energy matrix and fermion operators satisfying the usual commutation relations: $\hat{T}_{x,y} \equiv K_{x,y} c_x^\dagger c_y$ and $K_{x,y} \equiv -t(\delta_{x,y-1} + \delta_{x,y+1}) $.  (Please note that Roman subscripts now refer to lattice coordinates of the tight-binding model.) The marginal perturbation, for non interacting fermions, is a weak back-scattering term of strength $t^\prime$ acting at a single link located at point $p$.

The lattice entanglement temperature $1/\beta_x$ is determined by computing the reduced density matrix for an $[-l, l]$ sized sub-region, which Kilch \cite{KlichTe} has shown to be of the form,
\be
\label{lattice_rdm}
\hat{\rho} = Z^{-1}e^{-\sum{\beta_x \hat{T}_{x,y}}}
\ee
Using the general scheme developed by Peschel \cite{Peschel_corr_fn}, it may be shown that 
\be
\label{beta_numerical}
\beta_x K_{x,y} = [\log(G^{-1} - I)]_{x,y}
\ee
where $G_{x,y} = \langle c^\dagger_x c_y \rangle$ is the one particle correlation function calculated in the sub-region $x,y \in [-l,l]$.  Elements of the right hand side of equation (\ref{beta_numerical}) that lie off the tri-diagonal structure of $K_{x,y}$ were checked for exponential decay away from the diagonal and discarded. Since the eigenvalues of $G$ are exponentially close to $0$ or $1$, sufficient precision must be kept to determine $\beta_x$.  Holographically, all eigenvalues are needed to resolve the lowest energy features deep in the AdS space, where $z \approx l$. 

However, some further justification is required to interpret these results.  In general for small central charge (in our case $c = 1$) or small $N$ SU($N$) CFTs, quantum corrections to the Ryu-Takayanagi (RT) formula are needed in the bulk to compute quantities other than the entanglement entropy.  For instance, according to the semiclassical RT formula, the mutual information of well separated regions approaches zero, and quantum corrections give the dominant non-vanishing contributions \cite{Maldacena_qcorr}.  Similarly, computation of the Renyi entropies in the gravity dual do not have a simple semiclassical geometric interpretation.  Since the density matrix captures the entire operator content of the theory, it is natural to question whether a semiclassical gravity computation (or perturbation thereof) can fully describe the density matrix of the boundary theory \cite{refB}.

The $c =1$ CFT used in our calculations can be realized as free fermions for (Luttinger parameter) $g = 1$ or compactified free bosons for arbitrary $g$. In a noninteracting theory, the reduced density matrix takes the simple quadratic second quantized form in equation (\ref{lattice_rdm}) and the semiclassical geodesic appearing in the RT construction of entropy simply encodes the entanglement temperature function, $\beta(x)$.  Specifically, the equation of the geodesic is \begin{equation}
z^2(x) = {1 \over \pi} l\beta(x)
\end{equation}

In this special case, the semiclassical physics of the bulk completely describes the boundary theory density matrix for a single subregion.  When the boundary theory is deformed by a marginal perturbation, we have shown numerically that the quadratic form of the density matrix is preserved, but with a perturbed entanglement temperature and entropy.  On the gravity side, the marginal perturbation sources the AdS space in the neighborhood of the RT geodesic, perturbing the entropy.  It is reasonable to expect these two computations to be comparable, and the fact that the central charge appears identically in both computations supports this conclusion.  In this way, for the particular case of $c = 1$, we expect perturbations to the bulk AdS metric in the vicinity of the minimal surface may be extracted from the boundary computation of the entanglement temperature.

In this manuscript we have broadened the thermodynamic analogy that presently links quantum information and gravitational physics by uncovering a generalization of the Clausius relation. It is shown that application of an external field, coupled to a marginal operator in the boundary field theory, effects a local change in the entanglement temperature.  The resulting entropy change is then given by an equation analogous to the Clausius relation (\ref{delta_entropy},\ref{Clausius}), with spatial energy correlations at zero physical temperature playing the role of heat capacity. 

By comparing this relation to the holographic expression for entropy, we have identified a local quantity in the boundary theory, $\delta \beta(x,\xi(z))$, that appears to corresponds to a local degree of freedom in the dual gravity theory, the perturbed bulk metric $h_{xx}(x,z)$. We are therefore able to construct perturbations in the bulk geometry from direct diagonalizations of the boundary quantum theory. For the particular quantum impurity model we study, there is a significant difference between the analytic structure of the metric determined computationally from the boundary theory---it is singular at the position of the defect---and the metric calculated by gravitational perturbation theory.  In future work, it will be important to study extended perturbations, as opposed to a localized one, in that our computations suggest dramatic perturbations to the bulk geometry. 

We note that equation (\ref{delta_entropy}) may be generalized to finite physical temperature by using the appropriate $\beta(x)$ from reference \cite{KlichTe} and crosses over to the thermodynamic Clausius relation (\ref{Clausius}) at temperatures larger than the corresponding inverse length scale (or mass scale in the case of a system that is gapped).  Lastly we add that $\delta \beta(x)$ might be calculated analytically from ${\rm tr}(\rho {\cal O}(x)) = \< {\cal O}(x)\>$ where the left hand side is expanded to 1st order in $\delta \beta$ (as described leading to equation (\ref{nasty_exp})) and the right hand side is an ordinary perturbative expansion to 2nd order in $t^\prime/t$ at zero physical temp.  This is actually the basis of Peschel's numerical scheme in \cite{Peschel_corr_fn}; however, we have only been successful in computing $\delta \beta(x)$ numerically.

We wish to point out that recent work has come to our attention that studies gravity duals of marginally deformed CFTs from a quantum information perspective \cite{Taka_QI_metric}. Our research was supported by Research Corporation CC6535 (GL) and Howard Hughes Medical Institute Scholar Program (BC). GL wishes to acknowledge many useful discussions with Benjamin Burrington, Adam Durst, Barry Friedman and Daniel Yee.


\begin{references}
\bibitem{IntComment} Based upon our numerical data the integrand of (\ref{deltaS_QM}),  $\delta \beta(x)/z^4(x)$, vanishes strongly at the boundary and is approximately antisymmetric about $x=p$. Therefore, we have estimated  (\ref{deltaS_QM}) by $\delta S \sim \int{\delta \beta(x)/ l^2 dx}$.
\end{references}
\end{document}